\documentstyle[aps,prb,preprint]{revtex}
\begin{document}
\title{The Influence of Gradient Corrections on Bulk and Surface\\
Properties of TiO$_2$ and SnO$_2$}
\author{J.~Goniakowski, J.M.~Holender, L.N.~Kantorovich and M.J.~Gillan}
\address{Physics Department, Keele University, Staffordshire ST5 5BG, U.K.}
\author{J.A.~White}
\address{TCM Group, Cavendish Laboratory, Cambridge University,
Cambridge CB3 0HE, U.K.}
\date{\today}
\maketitle
\begin{abstract}
First-principles calculations based on density functional theory and
the pseudo\-potential method have been used to investigate the
influence of gradient corrections to the standard LDA technique
on the equilibrium structure and energetics of rutile TiO$_2$ and
SnO$_2$ perfect crystals and their (110) surfaces. We find that
gradient corrections increase the calculated lattice parameters by roughly
3~\%, as has been found for other types of material. Gradient
corrections give only very minor changes to the equilibrium surface
structure, but reduce the surface energies by about 30~\%.
\end{abstract}

\section{Introduction}
Most first-principles calculations on condensed matter are nowadays based
on density functional theory (DFT)~\cite{hoh64,koh65,jon89,gil91,pay92}.
This theory is formally exact,
but in practice an approximation has to be made to the exchange-correlation
energy, and the vast majority of calculations employ the local density
approximation (LDA). The basic assumption is that the exchange-correlation
energy per electron at any point in the system is related to the electron
density at that point in the same way as in a uniform electron gas,
and density gradients are ignored. Some theoretical justification
can be given for this~\cite{har74,gun76,lan77},
and in practice the LDA works well
in a wide range of situations. However, its accuracy is not always
satisfactory, particularly when energy differences associated with
changes of bonding are needed, as in e.g. molecular dissociation or
the adsorption of molecules at surfaces. Attempts to improve the
situation by adding lowest-order corrections in powers of the density
gradient are not successful, but important progress has been made
recently by requiring that the dependence of the energy on the gradients
satisfies certain physical requirements. This has led to various
forms of generalized gradient corrections
(GGC)~\cite{lan83,per86,pw86,bec88,lac93}.

In the last few years, there has been a large amount of work on the
influence of different GGC schemes on the total energies of
atoms and molecules~\cite{lan83,per86,pw86,lan85,kut88,bos90,mly91},
the equilibrium structure and cohesive energies
of covalent crystals~\cite{gar92,kon90,ori92,jua93,sei95},
the ground state of iron~\cite{zhu92,hag93},
and the energetics
of molecular adsorption on metal surfaces~\cite{whi94a,hu94,gun94}.
However, so far as we are
aware, there has been little work on the effect of GGC on the
properties of partially ionic materials such as the oxides TiO$_2$ and
SnO$_2$ treated here. The surface properties of materials like these
are extremely important, because of their application as gas sensors and
catalysts. We have recently reported a detailed study of the bulk and surface
properties of SnO$_2$~\cite{IanYY,IanXX},
and we have initiated work on the interaction
of molecules with the surfaces of both TiO$_2$ and SnO$_2$~\cite{unp}.
An understanding
of GGC is of considerable importance in this general area. The goal
of the present paper is to study the effect on the bulk and surface
properties of TiO$_2$ and SnO$_2$ of the two widely used GGC schemes due to
Perdew and Wang~\cite{per86,pw86}
and Becke and Perdew~\cite{per86,bec88}.

\section{Techniques}
The calculations are performed using the pseudopotential
method~\cite{gil91,pay92},
so that only the valence electrons are represented explicitly, the
valence-core interaction being represented by non-local norm-conserving
pseudopotentials, which are generated by first-principles calculations
on isolated atoms. Periodic boundary conditions are used, with the occupied
electronic orbitals expanded in a plane-wave basis. The expansion includes
all plane waves whose kinetic energy $\hbar^2 k^2 / 2m$ ($k$ the
wavevector, $m$ the electronic mass) is less than a chosen cutoff energy
$E_{\rm cut}$. The inclusions of gradient corrections within the
pseudopotential plane-wave technique has recently been discussed
in detail by White and Bird~\cite{whi94}, who show that a robust and
accurate calculation of the GGC exchange-correlation energy and potential
can be achieved by summation on exactly the same real-space grid as
would be used for the LDA. This technique has been used in the present work.

The first-principles pseudopotentials in Kleinman-Bylander
representation~\cite{kle82} were generated using the optimization
scheme of Lin {\em et al.}~\cite{lin93} in order to reduce the required
value of the plane-wave cutoff $E_{\rm cut}$. The pseudopotentials
used in the GGC calculations were constructed consistently by including
gradient corrections in the generation scheme. The Sn
pseudopotential was generated using the $5s^2 5p^2$ configuration for
$s$- and $p$-wave components, and the $5s^15p^{0.5}5d^{0.5}$
configuration for the $d$-wave. The core radii were equal to 2.1, 2.1 and
2.5~a.u. for the $s$, $p$ and $d$ components respectively.
The Ti pseudopotential was generated using
the $4s^{1.85}3d^2$ configuration for $s$ and $d$ waves
and the $4s^14p^{0.5}3d^{0.5}$ configuration for the
$p$ wave, with core radii of 2.2, 1.5 and 2.4 a.u. for $s$, $p$ and $d$ waves
respectively.
The oxygen pseudopotential used in our LDA calculations
was generated
using the $2s^2 2p^4$ configuration for the $s$ and $p$ waves and the
$2s^2 2p^{2.5} 3d^{0.5}$ configuration for the $d$ wave,
with a single core radius of 1.65~a.u. For the
gradient-corrected oxygen pseudopotential, we have used the single
configuration $2s^2 2p^{3.5} 3d^{0.45}$ and the same core radius.
The use of a core radius of 1.65~a.u. means that there is an appreciable
overlap of the oxygen and metal core spheres in the SnO$_2$ and TiO$_2$
crystals, and in principle this could cause inaccuracies. However,
direct comparisons of the present results with our earlier work
on SnO$_2$~\cite{IanXX}, which employed an oxygen pseudopotential with the
smaller core radius of 1.25~a.u., show that any errors due to core overlap are
very small.
The calculations have been done using a plane wave cut-off
$E_{\rm cut}$ of 600~eV for SnO$_2$ and 1000~eV
for TiO$_2$. Our tests show that with these cut-offs the energy per
unit cell is converged to within 0.2~eV, the convergence with respect to
$E_{\rm cut}$ being not noticeably influenced by the inclusion of gradient
corrections, even though the gradient corrected pseudopotentials
are less smooth and regular than the LDA ones~\cite{gar92,ort91}.

The calculations were performed using the CETEP code~\cite{cla92}
(the parallel version of the serial CASTEP code~\cite{pay92}) running
on the 64-node Intel iPSC/860 machine at Daresbury Laboratory. The code
uses the band-by-band conjugate-gradient technique to minimize the total
energy with respect to plane-wave coefficients. The LDA calculations
were performed using the Ceperley-Alder (CA)
exchange-correlation function~\cite{cep80}.

For the ground states calculations Brillouin zone sampling is performed
using the lowest order Monkhorst-Pack set of  k--points~\cite{mon76},
as in our earlier work on SnO$_2$~\cite{IanYY}.
Electronic densities
of states (DOS) associated with the ground state were calculated
using the tetrahedron method~\cite{jep71,leh72}, with $k$-point
sampling corresponding to 750 tetrahedra in the whole Brillouin zone.

\section{Results and discussion}
\subsection{Perfect SnO$_2$ and TiO$_2$ crystals}

The 6-atom rutile unit cell of SnO$_2$ and TiO$_2$ is
characterized by the two lattice parameters $a$ and $c$ and the internal
parameter $u$: the positions of the four oxygens are $(\pm u, \pm u, 0)$,
$( \frac{1}{2} \pm u, \frac{1}{2} \mp u, \frac{1}{2} )$.
The equilibrium structure has then been determined by relaxation with
respect to the lattice parameters $a$ and $c$ and the
internal parameter $u$. The equilibrium values of these
parameters both with and without gradient corrections are given in
Table~\ref{tab1}.

As usually happens, there is a tendency for the LDA to underestimate the
lattice parameter. This is especially noticeable for SnO$_2$, where there
may also be an effect due to our treatment of the 4$d$ shell as
part of the core. The inclusion of gradient corrections tends to increase
the lattice parameters, as has already been found for semiconducting
and metallic systems~\cite{gar92,jua93,sei95}. The increase is 4~\%
or more for the Perdew-Wang GGC, and leads to results for $a$ and $c$ that
are appreciably greater than experimental values. For the Becke-Perdew
GGC, the increase is roughly 3~\%. Both the $c/a$ ratio and the $u$
parameter are almost unaffected, and this suggests that the gradient
corrections have the effect of an isotropic negative pressure, as
pointed out by Seifert {\em et al.}~\cite{sei95}.

We have calculated the electronic DOS for the SnO$_2$ perfect crystal using
both LDA and the two GGC schemes, but the changes caused by GGC are very
small and we do not show the results here.

\subsection{The SnO$_2$ and TiO$_2$~(110) surfaces}
Our calculations on the stoichiometric (110) surface of
both materials have been done with the usual repeating slab geometry.
The rutile structure can be regarded as consisting of (110) planes of atoms
containing both metal (M) and oxygen (O) atoms, separated by planes
containing oxygen alone, so that the sequence of planes is
O - M$_2$O$_2$ - O - O - M$_2$O$_2$ - O etc. The entire crystal can
then be built up of symmetrical 3-plane O - M$_2$O$_2$ - O units.
The slabs we use contain three of these units, and our repeating cell
contains 18 atoms (6~M and 12~O).
The perfect (110) surface consists of rows of bridging oxygens lying
above a metal-oxygen layer.
The vacuum separating the slabs has been taken wide enough to
ensure that interactions between neighboring slabs are small.
The width we use corresponds to two O - M$_2$O$_2$ - O units, and
is such that planes of bridging oxygens on
the surfaces facing each other across the vacuum are
separated by about 6.8~\AA.

The surface structure has been determined by relaxing the
entire system to equilibrium, and the calculations have
been done with and without gradient corrections. As in
our previous work on SnO$_2$~(110)~\cite{IanXX}, and the work of
Ramamoorthy {\em et al.} on TiO$_2$~(110)~\cite{ram94a}, we find
displacements of the surface atoms of order 0.1~\AA, with
5-fold and 6-fold coordinated metal atoms (M$_{\rm II}$ and M$_{\rm I}$) moving
respectively into and
out of the surface, in-plane oxygens (O$_{\rm II}$) moving out and bridging
oxygens (O$_{\rm I}$) moving very little. The changes of the bond lengths
between the surface atoms, including sub-bridging oxygens (O$_{\rm III}$)
and the uppermost oxygens (O$_{\rm IV}$) of the following O - M$_2$O$_2$ - O
unit,
for LDA and gradient corrected calculations are given in Table~\ref{tab2}.
{}From these results, it is clear that gradient corrections have
only a minor effect on the relaxed equilibrium structure.
As we have already noted for the perfect crystal case,
modifications of atomic structure with respect to LDA results are more
pronounced in the PW scheme.

We have calculated the surface formation energy in the standard way,
by subtracting
from the slab total energy (18 atoms) three times the energy of a 6-atom
perfect crystal unit cell and dividing by the total surface area.
We find that the relaxed surface energy of SnO$_2$~(110) is
1.66~Jm$^{-2}$ in the LDA, 1.13~Jm$^{-2}$ in PW-GGC and 1.16~Jm$^{-2}$ for
BP-GGC. The LDA result is close to the value of 1.50~Jm$^{-2}$ reported
earlier~\cite{IanXX}. For TiO$_2$~(110), the values are 1.14, 0.82
and 0.84~Jm$^{-2}$ respectively. Our LDA result
for TiO$_2$ (110) is close to the value of 1.06~Jm$^{-2}$ reported by
Ramamoorthy {\em et al.}~\cite{ram94a}.
Comparison of LDA and GGC results shows that gradient
corrections have a substantial effect on the surface energies.
For both GGC schemes, the surface energies are lowered by about 30\% with
respect to the LDA values, the difference between PW and BP being very small.
This decrease of surface energy by GGC is consistent with the general tendency
of gradient corrections to remove the systematic overestimation of
electronic binding energy in the LDA.

The electronic DOS of the SnO$_2$~(110) surface using the LDA and the two
GGC schemes are compared in Fig.~\ref{fig1}.
In order to separate out effects of electronic structure, all the calculations
are done at the equilibrium lattice parameters and the relaxed positions
produced by the BP scheme. Overall, the differences between the three sets
of results are small. However, there are significant differences at the top
of the valence band and at the top of the O(2s) band. As we found in our
previous work~\cite{IanYY} there are peaks at the top of both bands due to
surface states, these states being concentrated on the bridging oxygens.
The effect of GGC is to reduce the intensity of the peak at the top of
O(2s) band. The effect on the intensity of the valence band peak is less
systematic, since BP increases it but PW decreases it.
The reason why these effects are interesting is that there appears to be no
experimental evidence for the surface--state peak at the top of the valence
band, so that the LDA predictions seems not to be consistent with experiment.
The present results suggest the possibility that this inconsistency may be
due to inaccurate treatament of exchange and correlation.

\section{Conclusions}
Our calculations show that gradient corrections increase the lattice
parameters of TiO$_2$ and SnO$_2$ by $\sim$~4~\% for the Perdew-Wang
scheme and $\sim$~3~\% for the Becke-Perdew scheme. These effects
are similar to those reported previously for metals and semiconductors.
For the surfaces we examined, gradient corrections have very little effect
on the relaxed surface structure, but the surface energies are
substantially reduced -- by $\sim$~30~\% in both the Perdew-Wang
and the Becke-Perdew schemes. The effects of gradient corrections on the
electronic DOS of SnO$_2$~(110) surface are very small, except at the top
of the O(2s) and O(2p) bands. The changes we find at the top of the O(2p)
band may be relevant to apparent inconsistences between calculated and
experimental results for the surface DOS in this region.

\section{Acknowledgments}
The work of JG is supported by EPSRC grant GR/J34842. The major
calculations were performed on the Intel iPSC/860 parallel computer
at Daresbury Laboratory, and we are grateful for a generous
allocation of time on the machine. Analysis of the results was
performed using local hardware funded by EPSRC grant GR/J36266.

\begin{figure}
\caption{The electronic DOS of the SnO$_2$~(110) surface for LDA (solid line),
PW--GGC (dotted line) and BP--GGC (dashed line). The calculations are made at
the equilibrium lattice parameters of the BP scheme.
For presentation purposes, we have broadened the calculated DOS
by Gaussians of width 0.5~eV.}
\label{fig1}
\end{figure}

\begin{table}
\caption{Comparison of theoretical and experimental values of lattice
parameters $a$ and $c$ and the internal coordinate $u$ of SnO$_2$
and TiO$_2$. The theoretical values are calculated using the Ceperley-Alder
form of LDA (CA), and the Perdew-Wang (PW) and Becke-Perdew (BP) forms of GGC.
Experimental values are from Ref.~40.}
\begin{tabular}{lldddddd}
&&\multicolumn{2}{c}{$a$ (\AA) }&\multicolumn{2}{c}{$c$ (\AA)}&  $c/a$ & $u$
\\ \hline
SnO$_2$ & CA & 4.645 & (-1.9\%) & 3.060 & (-4.0\%) & 0.659 & 0.307 \\
        & PW & 4.868 &  (2.8\%) & 3.183 & ( 0.0\%) & 0.654 & 0.307 \\
        & BP & 4.809 &  (1.5\%) & 3.159 & (-0.8\%) & 0.657 & 0.307 \\
      & expt.& 4.737 &          & 3.186 &          & 0.673 & 0.307 \\ \hline
TiO$_2$ & CA & 4.625 &  (0.7\%) & 2.911 & (-1.6\%) & 0.629 & 0.305 \\
        & PW & 4.781 &  (4.1\%) & 3.072 &  (3.9\%) & 0.643 & 0.305 \\
        & BP & 4.747 &  (3.3\%) & 3.039 &  (2.7\%) & 0.640 & 0.305 \\
       &expt.& 4.594 &          & 2.958 &          & 0.644 & 0.305 \\
\end{tabular}
\label{tab1}
\end{table}

\begin{table}
\caption{Comparison of calculated bond length modifications on SnO$_2$~(110)
and TiO$_2$~(110) with respect to the bulk values for LDA (CA),
and two alternative forms of GGC exchange-correlation.}
\begin{tabular}{ldddddd}
&\multicolumn{3}{c}{SnO$_2$}&\multicolumn{3}{c}{TiO$_2$}  \\ \hline
       & CA      & PW   & BP      & CA      & PW      & BP        \\
O$_{\rm IV}$  -- M$_{\rm II}$ &  -4.5\% &  -4.9\%  &  -4.2\% &  -5.6\% &
-5.5\% &  -5.6\%  \\
O$_{\rm I}$   -- M$_{\rm I}$  &  -3.7\% &  -4.0\%  &  -3.8\% &  -4.9\% &
-5.5\% &  -5.5\%  \\
O$_{\rm II}$  -- M$_{\rm II}$ &  -1.2\% &  -1.2\%  &  -1.2\% &  -0.9\% &
-1.2\% &  -1.2\%  \\
O$_{\rm III}$ -- M$_{\rm I}$  &   4.3\% &   4.3\%  &   4.8\% &   4.6\% &
4.7\% &   4.5\%  \\
O$_{\rm II}$  -- M$_{\rm I}$  &   2.7\% &   2.9\%  &   2.9\% &   2.3\% &
2.8\% &   2.8\%  \\
\end{tabular}
\label{tab2}
\end{table}
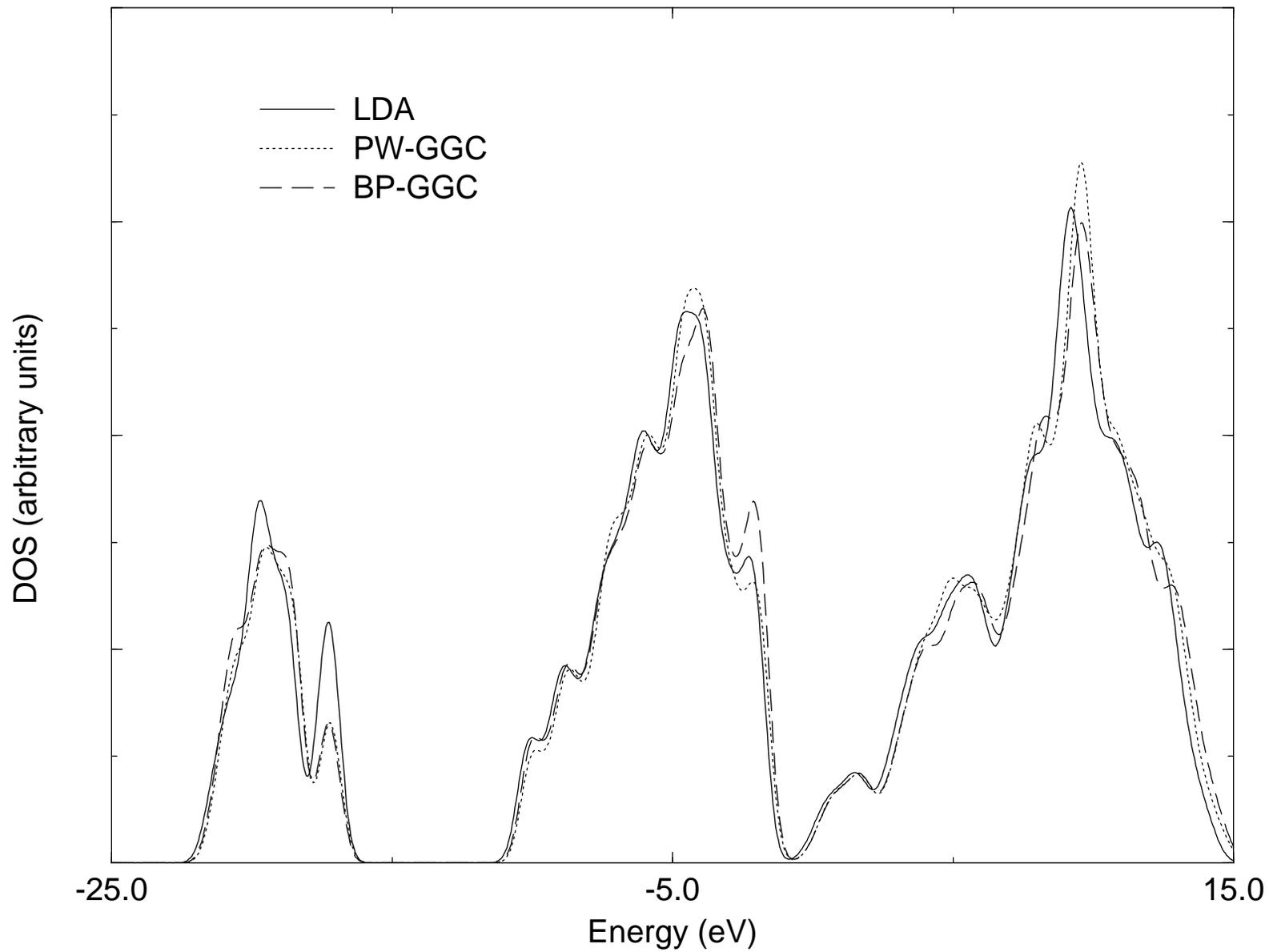
\end{document}